\documentstyle[12pt]{article}

\begin{document}

\begin{flushright}
DPUR/TH/69\\
June, 2020\\
\end{flushright}
\vspace{20pt}

\pagestyle{empty}
\baselineskip15pt

\begin{center}
{\large\bf  Higgs Potential from Weyl Conformal Gravity
\vskip 1mm }

\vspace{20mm}

Ichiro Oda\footnote{
           E-mail address:\ ioda@sci.u-ryukyu.ac.jp
                  }

\vspace{10mm}
           Department of Physics, Faculty of Science, University of the 
           Ryukyus,\\
           Nishihara, Okinawa 903-0213, Japan\\

\end{center}


\vspace{10mm}
\begin{abstract}

We consider Weyl's conformal gravity coupled to a complex matter field in Weyl geometry.
It is shown that a Higgs potential naturally arises from a $\tilde R^2$ term in moving 
from the Jordan frame to the Einstein frame. A massless Nambu-Goldstone boson, which 
stems from spontaneous symmetry breakdown of the Weyl gauge invariance, is absorbed 
into the Weyl gauge field, thereby the gauge field becoming massive. 
We present a model where the gravitational interaction generates a Higgs potential whose 
form is a perfect square.
Finally, we show that a theory in the Jordan frame is gauge-equivalent to the corresponding
theory in the Einstein frame via the BRST formalism. 
\end{abstract}

\newpage
\pagestyle{plain}
\pagenumbering{arabic}


\section{Introduction}

One of the most important purposes of modern particle physics is to find symmetries hidden
in nature and study their breaking mechanism.  It is somewhat surprising that there are not
so many symmetries which are useful in clarifying the fundamental laws existing in nature
owing to Coleman and Mandula's no-go theorem \cite{Coleman} which states that no Lorentz 
non-scalar charges other than Poincare generators can be S-matrix symmetries in Poincare-invariant
theories.
 
As one of such important symmetries which are not against the Coleman-Mandula theorem,
we have a conformal symmetry. In particular, a scale symmetry in the conformal symmetry
is a mysterious symmetry and occupies the special position in the sense that it always appears 
as an approximate symmetry even if it is ubiquitous from particle physics to cosmology \cite{Coleman2}. 
The reason is that the scale symmetry prohibits definite mass or length scales but nature is full of various 
kinds of scales and consequently the scale symmetry is broken spontaneously or explicitly. Understanding 
of the origin of different mass scales could be a key step toward the resolution of the hierarchy problems 
such as the gauge hierarchy problem and the elusive cosmological constant problem. 

In this article, we would like to study the theory with a global or a local scale symmetry,\footnote{We sometimes
refer to either a global or a local scale symmetry as a scale symmetry or a Weyl symmetry in this article.}
especially from the viewpoint of its spontaneous symmetry breakdown and emergence of the Higgs
potential. As is well-known, a natural avenue of developments in quantum field theories is to extend 
a global scale symmetry to a local Weyl symmetry. This is in particular true of gravitational theories 
owing to no-hair theorems of black holes \cite{MTW}. After a painful retreat due to the second clock 
problem \cite{Penrose}, we have recently watched a revival of interests in Weyl's conformal gravity 
\cite{Weyl}-\cite{Tang} since this gravitational theory provides us with a playground for not only 
treating the Weyl symmetry but also supplying us with a candidate of dark matter, which is a massive 
Weyl gauge field.      

This article is organized as follows: In Section 2, we review a Higgs potential emerging from an $R^2$ term 
in theories with a restricted Weyl symmetry \cite{Edery1}-\cite{Oda-New} and a global scale symmetry 
and point out its problems. In Section 3, we show that a massless dilaton is absorbed into a Weyl gauge 
field and therefore the problems of cosmology \cite{Witten, Nishino} and the fifth force \cite{Will}, 
which are associated with the massless dilaton, are solved in Weyl's conformal gravity. 
Furthermore, we present a new model where a Higgs potential arises from the gravity and discuss 
how we can obtain the electroweak scale from the Planck scale by selecting the parameters belonging to 
the gravitational sector.  A peculiar feature of this Higgs potential is that it has the form of a perfect square
so the cosmological constant at the minimum is identically vanishing.
In Section 4, we rederive the Lagrangian density obtained by a change of variables in Section 3 through 
the BRST formalism, and show that the Lagrangian density in the Jordan frame is gauge-equivalent to
that in the Einstein frame. The final section is devoted to conclusion.

\section{Review of Higgs potential from $R^2$ term}

We begin with a review of emergence of a Higgs potential from an $R^2$ term in a gravitational theory coupled to 
a $U(1)$ gauge theory with a complex scalar field whose Lagrangian density is given by \cite{Oda-New}$\footnote{We 
follow the conventions and notation of the MTW textbook \cite{MTW}.}$:
\begin{eqnarray}
{\cal{L}} = \sqrt{-g} \left( \xi_1^2 R^2 + \xi_2 R |\Phi|^2 - |D_\mu \Phi|^2 - \lambda |\Phi|^4 
- \frac{1}{4} F_{\mu\nu} F^{\mu\nu} \right),
\label{Weyl Lag0}  
\end{eqnarray}
where $\xi_1, \xi_2, \lambda$ are dimensionless coupling constants, $A_\mu, F_{\mu\nu}, \Phi, D_\mu \Phi$ 
are respectively a $U(1)$ gauge field, its field strength defined as $F_{\mu\nu} \equiv \partial_\mu A_\nu
- \partial_\nu A_\mu$, a complex scalar field, and its covariant derivative defined as 
$D_\mu \Phi \equiv ( \partial_\mu - i e A_\mu ) \Phi$. The parameter $\xi_1^2$ is positive to avoid 
tachyons \cite{Stelle} and the gravitational coupling constant corresponds to $\frac{1}{\xi_1}$.
Finally, note that the complex scalar field couples to the scalar curvature via a nonminimal coupling
term $\xi_2 R |\Phi|^2$.

For simplicity of writing, in this section we drop the gauge field $A_\mu$ and we work with the
following Lagrangian density:
\begin{eqnarray}
{\cal{L}} = \sqrt{-g} \left( \xi_1^2 R^2 + \xi_2 R |\Phi|^2 - |\partial_\mu \Phi|^2 
- \lambda |\Phi|^4 \right).
\label{Weyl Lag-J}  
\end{eqnarray}

This Lagrangian density is invariant under both a global scale transformation ($\Omega$ = constant) and 
a restricted Weyl transformation  
\begin{eqnarray}
g_{\mu\nu} \rightarrow g^\prime_{\mu\nu} = \Omega^2 (x) g_{\mu\nu}, \qquad 
\Phi \rightarrow \Phi^\prime = \Omega^{-1}(x) \Phi, 
\label{Res-Weyl}  
\end{eqnarray}
where the gauge parameter obeys a constraint $\Box \Omega = 0$ \cite{Edery1}-\cite{Oda-New}. 
In order to prove the restricted Weyl invariance, we need to use the following transformation of the scalar 
curvature under (\ref{Res-Weyl}):
\begin{eqnarray}
R \rightarrow R^\prime = \Omega^{-2} ( R - 6 \Omega^{-1} \Box \Omega ). 
\label{Weyl-R}  
\end{eqnarray}

Now we are ready to show that a Higgs potential emerges for the Higgs field $\Phi$ in addition to a scale-invariant 
potential term $\lambda |\Phi|^4$ in (\ref{Weyl Lag-J}).
The first key observation is that the $\xi_1^2 R^2$ term can be cast to the form of the scalar-tensor
gravity, $\varphi R - \frac{1}{4 \xi_1^2} \varphi^2$ where $\varphi$ is a scalar field with the dimension 
of mass squared. Thus, the Lagrangian density (\ref{Weyl Lag-J}) reads
\begin{eqnarray}
{\cal{L}} = \sqrt{-g} \left( \varphi R - \frac{1}{4 \xi_1^2} \varphi^2 + \xi_2 R |\Phi|^2 - |\partial_\mu \Phi|^2 
- \lambda |\Phi|^4 \right).
\label{Weyl Lag2}  
\end{eqnarray} 
Next, let us move from the Jordan frame to the Einstein frame. To do so, we will do a change of variables,
which takes the same form as a local conformal transformation\footnote{In the conventional approach, we consider a
conformal transformation of only the metric but it usually yields non-canonical kinetic terms and
non-polynomial potentials. To avoid such a situation, we also consider a conformal transformation of 
the matter field.} 
\begin{eqnarray}
g_{\mu\nu} \rightarrow g_{\ast\mu\nu} = \Omega^2 (x) g_{\mu\nu}, \qquad 
\Phi \rightarrow \Phi_\ast = \Omega^{-1}(x) \Phi,
\label{Transf-E-frame}  
\end{eqnarray}
except the scalar field $\varphi$.\footnote{As can be understood in the next model, we could also consider 
the conformal transformation of $\varphi$, $\varphi \rightarrow \varphi_\ast = \Omega^{-2}(x) \varphi$, 
without changing the final result, but the method in hand is more useful in seeing the role of a conformal 
factor.} Henceforth, we express quantities in the Einstein frame by putting the symbol $\ast$ on them.

Under this transformation (\ref{Transf-E-frame}) we have formulae \cite{Fujii}
\begin{eqnarray}
\sqrt{-g} = \Omega^{-4} \sqrt{-g_\ast}, \qquad
R = \Omega^2 ( R_\ast + 6 \Box_\ast F - 6 g_\ast^{\mu\nu} F_\mu F_\nu ), 
\label{Transf-E-frame2}  
\end{eqnarray}
where we have defined
\begin{eqnarray}
F \equiv \log \Omega, \quad
\Box_\ast F \equiv \frac{1}{\sqrt{- g_\ast}} \partial_\mu ( \sqrt{- g_\ast} g_\ast^{\mu\nu} \partial_\nu F), \quad
F_\mu \equiv \partial_\mu F = \frac{\partial_\mu \Omega}{\Omega}.
\label{F}  
\end{eqnarray}

Then, the Lagrangian density (\ref{Weyl Lag2}) is rewritten as
\begin{eqnarray}
{\cal{L}} &=& \sqrt{-g_\ast} \Biggl[ ( \varphi \Omega^{-2} + \xi_2 |\Phi_\ast|^2 ) 
( R_\ast + 6 \Box_\ast F - 6 g_\ast^{\mu\nu} F_\mu F_\nu ) - \frac{1}{4 \xi_1^2} \varphi^2 \Omega^{-4}  
\nonumber\\
&-& \Omega^{-2} g_\ast^{\mu\nu} \partial_\mu (\Omega \Phi_\ast^\dagger) \partial_\nu 
(\Omega \Phi_\ast) - \lambda |\Phi_\ast|^4  \Biggr].
\label{E-Lag}  
\end{eqnarray}
To reach the Einstein frame, we have to choose a conformal factor $\Omega(x)$ to satisfy 
a relation
\begin{eqnarray}
\varphi \Omega^{-2} = - \xi_2 |\Phi_\ast|^2 + \frac{M_{Pl}^2}{2},
\label{Planck-mass}  
\end{eqnarray}
where $M_{Pl}$ is the reduced Planck mass. As a result, with the redefinition $\omega(x) \equiv \sqrt{6}
M_{Pl} F(x)$,\footnote{This redefinition implies that we use $\omega(x)$ instead of $\varphi(x)$ as a dynamical
degree of freedom.} we obtain a Lagrangian density in the Einstein frame:
\begin{eqnarray}
{\cal{L}} &=& \sqrt{-g_\ast} \Biggl[ \frac{M_{Pl}^2}{2} R_\ast - \frac{1}{2} g_\ast^{\mu\nu} 
\partial_\mu \omega \partial_\nu \omega - |\partial_\mu \Phi_\ast|^2
- \frac{1}{16 \xi_1^2} M_{Pl}^4 + \frac{\xi_2}{4 \xi_1^2} M_{Pl}^2 |\Phi_\ast|^2
\nonumber\\
&-& \left( \lambda + \frac{\xi_2^2}{4 \xi_1^2} \right) |\Phi_\ast|^4 
+ \left( \frac{1}{\sqrt{6} M_{Pl}} \Box_\ast \omega - \frac{1}{6 M_{Pl}^2} g_\ast^{\mu\nu} 
\partial_\mu \omega \partial_\nu \omega \right) |\Phi_\ast|^2 \Biggr].
\label{E-Lag2}  
\end{eqnarray}
It is worthwhile to notice that spontaneous symmetry breakdown of a scale invariance
has occurred and consequently we have a massless Nambu-Goldstone boson $\omega(x)$, which is
often called ``dilaton''.  Note that the kinetic term for the dilaton comes from 
$g_\ast^{\mu\nu} F_\mu F_\nu$ in Eq. (\ref{E-Lag}). 

Let us now explain the reason why the spontaneous symmetry breakdown of a scale invariance
has occurred in this model. (This reasoning can be also applied for the other models considered in this article
with a suitable modification.) From the Lagrangian density (\ref{Weyl Lag2}), the full potential of
the two scalar fields is given by
\begin{eqnarray}
V(\varphi, |\Phi|) = \sqrt{-g} \left( - \varphi R + \frac{1}{4 \xi_1^2} \varphi^2 - \xi_2 R |\Phi|^2 
+ \lambda |\Phi|^4 \right).
\label{Full-pot}  
\end{eqnarray} 
In order to understand a common ground state of the two scalar fields, we have to find (local) minima
of the potential (\ref{Full-pot}) in the both variables. For the minima, the gradient of the potential,
$(\partial_\varphi V, \partial_{|\Phi|} V)$, must vanish. The solution is given by
\begin{eqnarray}
\langle |\Phi|^2 \rangle = \frac{\xi_2}{4 \lambda \xi_1^2} \langle \varphi \rangle, \qquad
\langle \varphi \rangle = 2 \xi_1^2 \langle R \rangle,
\label{Pot-minimum}  
\end{eqnarray} 
where $\langle A \rangle$ denotes a vacuum expectation value for a generic field $A$.
We can also verify that these extremal values are indeed local minima by evaluating the Hessian provided that 
$\xi_2 \langle R \rangle$ is positive. In the above theory, among the degenerate minima, we have chosen 
a specific configuration given by
\begin{eqnarray}
\langle \varphi \rangle = \frac{2 \lambda \xi_1^2}{\xi_2^2 + 4 \lambda \xi_1^2} M_{Pl}^2,
\label{A minimum}  
\end{eqnarray} 
by which the spontaneous symmetry breakdown of a scale symmetry has been triggered. Note that
since in this theory the vacuum expectation value of the scalar curvature is given by
\begin{eqnarray}
\langle R \rangle = \frac{\lambda}{\xi_2^2 + 4 \lambda \xi_1^2} M_{Pl}^2,
\label{VEV-R}  
\end{eqnarray} 
it is positive definite due to $\lambda > 0$.

A remarkable thing in the theory under consideration is that a Higgs potential is generated from
the $R^2$ term together with the $R |\Phi|^2$. Actually, a gauge symmetry is spontaneously 
broken if we choose the parameters to be 
\begin{eqnarray}
\frac{\xi_2}{4 \xi_1^2} > 0, \qquad \lambda + \frac{\xi_2^2}{4 \xi_1^2} > 0.
\label{Parameter}  
\end{eqnarray}

Of course, we can construct similar models where spontaneous symmetry breakdown of a scale symmetry 
yields a Higgs potential. For comparison, let us present such a model whose Lagrangian
density is composed of a scalar-tensor gravity with scale-invariant potentials made out of
a real scalar field $\phi$ and a complex scalar field $\Phi$. To simplify the argument, we consider
the case where only the real scalar $\phi$ couples to a scalar curvature as follows \cite{Oda-Planck, Oda-C}:
\begin{eqnarray}
{\cal{L}} = \sqrt{-g} \Biggl[ \frac{1}{2} \xi \phi^2 R - \frac{1}{2} ( \partial_\mu \phi )^2 - |\partial_\mu \Phi|^2 
- \lambda_1 \phi^4 - \lambda_2 \phi^2 |\Phi|^2 - \lambda_3 |\Phi|^4 \Biggr],
\label{ST Lag-J}  
\end{eqnarray}
where we assume $\xi \neq - \frac{1}{6}$ to avoid the case of a local scale invariance, and $\lambda_i ( i = 1, 2, 3 )$
are dimensionless coupling constants.
Following the same line of the argument as before, we do a change of variables in Eq. (\ref{Transf-E-frame})
as well as $\phi \rightarrow \phi_\ast = \Omega^{-1}(x) \phi$. Moving to the Einstein frame requires us to
choose a conformal factor $\Omega (x)$ to be
\begin{eqnarray}
\xi \phi^2 = \Omega^2 M_{Pl}^2,
\label{E-frame-cond1}  
\end{eqnarray} 
or equivalently,   
\begin{eqnarray}
\xi \phi_\ast^2 = M_{Pl}^2.
\label{E-frame-cond2}  
\end{eqnarray} 
Consequently, we can obtain the final expression
\begin{eqnarray}
{\cal{L}} &=& \sqrt{-g_\ast} \Biggl[ \frac{M_{Pl}^2}{2} R_\ast - \frac{1}{2} g_\ast^{\mu\nu} 
\partial_\mu \hat \omega \partial_\nu \hat \omega - |{\cal{D}}_\mu \Phi_\ast|^2
- \frac{\lambda_1}{\xi^2} M_{Pl}^4
\nonumber\\
&-& \frac{\lambda_2}{\xi} M_{Pl}^2 |\Phi_\ast|^2 
- \lambda_3 |\Phi_\ast|^4 \Biggr],
\label{ST Lag-E}  
\end{eqnarray}
where we have defined 
\begin{eqnarray}
\hat \omega \equiv \sqrt{\frac{6 \xi + 1}{\xi}} M_{Pl} F, \qquad
{\cal{D}}_\mu \Phi_\ast \equiv \left( \partial_\mu + \sqrt{\frac{\xi}{6 \xi + 1}} \frac{1}{M_{Pl}}
\partial_\mu \hat \omega \right) \Phi_\ast.
\label{Def-Dphi}  
\end{eqnarray}     

Let us note that even in this simple model a Higgs potential emerges when a scale symmetry is
spontaneously broken and as a result a Nambu-Goldstone boson $\hat \omega$ appears in the mass
spectrum. However, there is a big difference: In (\ref{E-Lag2}), the Higgs potential arises from an $R^2$ 
term with the help of the nonminimal term while in  (\ref{ST Lag-E}) it comes from the scale-invariant 
potentials which already existed in the classical action. In this sense, we can state that the Higgs potential
in (\ref{E-Lag2}) is of the gravitational origin. 

To close this section, we should pick up two issues which will be clarified in Sections 3 and 4. 
First, it is known that the presence of a massless dilaton causes cosmological problems through the gravity
at large scales since the dilaton couples to any fields in a universal manner \cite{Witten, Nishino}. 
In addition, if the massless dilaton couples to the matter directly at the level of an action, the weak 
equivalence principle is violated, thereby yielding the fifth force, but at present there is no such a force 
by experiments in the solar system \cite{Will}. 

Thus, the dilaton should either have a mass anyway or be absorbed into a gauge field. 
Indeed, the dilaton could acquire a small mass via trace anomaly \cite{Fujii}. In the next section, we will
pursue an alternative possibility that the massless dilaton is absorbed into a Weyl gauge field, thereby 
the Weyl gauge field becoming massive and at the same the massless dilaton disappearing from the mass spectrum.
Moreover, we will present a model where a Higgs potential whose form is a perfect square, is generated by the
gravitational interaction. 

As a second issue, there is an ongoing debate with a long history on the equivalence between the Jordan 
frame and the Einstein frame \cite{Kamenshchik}. In the derivation of the Higgs potentials, we have heavily 
relied on the equivalence between the Jordan frame and the Einstein frame, so it would be more desirable 
if we could derive the Higgs potentials via a different but more sound method. To this end,  in Section 4
we will use the BRST formalism and prove the quantum equivalence of the theory between the Jordan frame
and the Einstein one. This proof is one of advantages in our formalism in the sense that the existence of 
a local scale invariance makes it possible to show a gauge-equivalence between the two frames.

\section{Weyl's conformal gravity and Higgs potential}

To remove a massless Nambu-Goldstone boson, i.e., the dilaton, from the mass spectrum, we make 
use of Weyl's conformal gravity where a global scale invariance is promoted to a local scale one.
When we consider a $U(1)$ gauge theory with a complex scalar field coupled to the gravity in Weyl
geometry, the most general Lagrangian density, which is invariant under Weyl gauge transformation (discussed
shortly), is given by\footnote{ We have used the conventions and notation in Ref. \cite{Oda-Planck} 
for the Weyl geometry.} 
\begin{eqnarray}
{\cal{L}} &=& \sqrt{-g} \Biggl( - \frac{1}{2 \xi_0^2} \tilde C_{\mu\nu\rho\sigma} \tilde C^{\mu\nu\rho\sigma} 
+ \xi_1^2 \tilde R^2 + \xi_2 \tilde R |\Phi|^2 - \frac{1}{4} H_{\mu\nu} H^{\mu\nu} 
\nonumber\\
&-& \frac{1}{4} F_{\mu\nu} F^{\mu\nu} - |D_\mu \Phi|^2 - \lambda |\Phi|^4  \Biggr),
\label{Weyl-U1-Lag}  
\end{eqnarray}
where $\tilde C_{\mu\nu\rho\sigma}, \tilde R$ are respectively the conformal tensor and scalar curvature
in the Weyl geometry. For instance, $\tilde R$ is defined as
\begin{eqnarray}
\tilde R \equiv g^{\mu\nu} \tilde R_{\mu\nu} 
= R - 6 f \nabla_\mu S^\mu - 6 f^2 S_\mu S^\mu,
\label{W-scalar-curv}
\end{eqnarray}
where $R, S_\mu$ and $f$ are the scalar curvature in the Riemann geometry, the Weyl gauge field and
the coupling constant for a noncompact Abelian group, respectively. Moreover, $H_{\mu\nu}$ is the field
strength of the Weyl gauge field and $D_\mu \Phi$ is a covariant derivative, which are defined as
\begin{eqnarray}
H_{\mu\nu} = \partial_\mu S_\nu - \partial_\nu S_\mu, \qquad
D_\mu \Phi = (  \partial_\mu - f S_\mu - i e A_\mu ) \Phi.
\label{H-DPhi}
\end{eqnarray}

For simplicity of writing, we will put $\xi_0^{-2} = A_\mu = 0$ and work with the following Langrangian
density:
\begin{eqnarray}
{\cal{L}} = \sqrt{-g} \Biggl(  \xi_1^2 \tilde R^2 + \xi_2 \tilde R |\Phi|^2 - \frac{1}{4} H_{\mu\nu} H^{\mu\nu} 
- |D_\mu \Phi|^2 - \lambda |\Phi|^4  \Biggr),
\label{Reduced-Lag}  
\end{eqnarray}
where we now have $D_\mu \Phi = (  \partial_\mu - f S_\mu ) \Phi$ since we have dropped the $U(1)$ gauge
field $A_\mu$ in Eq. (\ref{H-DPhi}).
The Weyl gauge transformation reads
\begin{eqnarray}
g_{\mu\nu} \rightarrow g^\prime_{\mu\nu} = \Omega^2 (x) g_{\mu\nu}, \;
\Phi \rightarrow \Phi^\prime = \Omega^{-1}(x) \Phi, \;
S_\mu \rightarrow S^\prime_\mu = S_\mu - \frac{1}{f} \partial_\mu \log \Omega. 
\label{Weyl gauge transf}  
\end{eqnarray}

The argument proceeds in the same fashion as that in Section 2. By introducing 
a scalar field $\varphi$, we can rewrite (\ref{Reduced-Lag}) into the form  
\begin{eqnarray}
{\cal{L}} = \sqrt{-g} \Biggl[ \Bigl( \varphi +   \xi_2 |\Phi|^2 \Bigr) \tilde R  - \frac{1}{4 \xi_1^2} \varphi^2
- \frac{1}{4} H_{\mu\nu} H^{\mu\nu} - |D_\mu \Phi|^2 - \lambda |\Phi|^4  \Biggr].
\label{varphi-Lag}  
\end{eqnarray}
To move from the Jordan frame to the Einstein frame, we will do a change of variables,
which is the Weyl gauge transformation except the scalar field $\varphi$:
\begin{eqnarray}
g_{\mu\nu} &\rightarrow& g_{\ast\mu\nu} = \Omega^2 (x) g_{\mu\nu}, \qquad 
\Phi \rightarrow \Phi_\ast = \Omega^{-1}(x) \Phi, 
\nonumber\\
S_\mu &\rightarrow& S_{\ast\mu} = S_\mu - \frac{1}{f} \partial_\mu \log \Omega
\equiv S_\mu - \frac{1}{f} \partial_\mu F.
\label{E-Weyl gauge transf}  
\end{eqnarray}
As a result, we find that
\begin{eqnarray}
{\cal{L}} &=& \sqrt{-g_\ast} \Biggl[ \Bigl( \varphi \Omega^{-2} +   \xi_2 |\Phi_\ast|^2 \Bigr) 
( R_\ast - 6 f g_\ast^{\mu\nu} \nabla_{\ast\mu} S_{\ast\nu} - 6 f^2 g_\ast^{\mu\nu} 
S_{\ast\mu} S_{\ast\nu} )
\nonumber\\
&-& \frac{1}{4 \xi_1^2} \varphi^2 \Omega^{-4} - \frac{1}{4} g_\ast^{\mu\nu} g_\ast^{\alpha\beta} 
H_{\mu\alpha} H_{\nu\beta} - g_\ast^{\mu\nu} D_{\ast\mu} \Phi_\ast^\dagger 
D_{\ast\nu} \Phi_\ast - \lambda |\Phi_\ast|^4  \Biggr],
\label{varphi-Lag-star}  
\end{eqnarray}
where we have defined
\begin{eqnarray}
\nabla_{\ast\mu} S_{\ast\nu} \equiv \partial_\mu S_{\ast\nu} - \Gamma_{\ast\mu\nu}^\lambda 
S_{\ast\lambda}, \qquad
D_{\ast\mu} \Phi_\ast \equiv ( \partial_\mu - f S_{\ast\mu} ) \Phi_\ast.
\label{Def-Star}  
\end{eqnarray}

Fixing a conformal factor as in (\ref{Planck-mass}) leads to the final expression 
\begin{eqnarray}
{\cal{L}} &=& \sqrt{-g_\ast} \Biggl[ \frac{M_{Pl}^2}{2} R_\ast - \frac{1}{2} m_S^2 g_\ast^{\mu\nu} 
S_{\ast\mu} S_{\ast\nu} - \frac{1}{4} g_\ast^{\mu\nu} g_\ast^{\alpha\beta} H_{\mu\alpha} H_{\nu\beta}
- |D_{\ast\mu} \Phi_\ast|^2 
\nonumber\\
&-& \lambda |\Phi_\ast|^4 - \frac{1}{4 \xi_1^2} \left( \xi_2 |\Phi_\ast|^2 
- \frac{M_{Pl}^2}{2} \right)^2  \Biggr],
\label{F-Lag-star}  
\end{eqnarray}
where the mass of the Weyl gauge field is $m_s = \sqrt{6} f M_{Pl}$ and $H_{\mu\nu} 
= \partial_\mu S_\nu - \partial_\nu S_\mu = \partial_\mu S_{\ast\nu} - \partial_\nu S_{\ast\mu}$.
Comparing with the case of a global scale symmetry, in this case $F_\mu$ is replaced with $S_\mu$, 
and as seen in (\ref{E-Weyl gauge transf}) the Nambu-Goldstone boson $F$ is absorbed
into the Weyl gauge field $S_{\ast\mu}$.
In this way, we have shown in an explicit manner that the Nambu-Goldstone boson, which
comes from spontaneous symmetry breakdown of the Weyl gauge symmetry, is absorbed into
the Weyl gauge field, thereby the Weyl gauge having the mass $m_S$. Since we have no
more a massless dilaton, we are now free from problems of both cosmology and the fifth force  
associated with the dilaton. 

Next, let us notice that after spontaneous symmetry breakdown of the Weyl gauge symmetry, not only the
massive Weyl gauge field but also a new Higgs potential have appeared in Eq. (\ref{F-Lag-star}). The Higgs 
potential can be read as
\begin{eqnarray}
V(\Phi_\ast) &=& \lambda |\Phi_\ast|^4 + \frac{1}{4 \xi_1^2} \Bigl( \xi_2 |\Phi_\ast|^2 - \frac{M_{Pl}^2}{2} \Bigr)^2
\nonumber\\
&=& \Bigl( \lambda + \frac{\xi_2^2}{4 \xi_1^2} \Bigr) \Biggl[ |\Phi_\ast|^2 
- \frac{\xi_2}{2 (\xi_2^2 + 4 \lambda \xi_1^2)} M_{Pl}^2 \Biggr]^2
\nonumber\\
&+& \frac{\lambda}{4 (\xi_2^2 + 4 \lambda \xi_1^2)} M_{Pl}^4.
\label{Higgs-pot}  
\end{eqnarray}   
Then, we obtain a minimum of the potential and a cosmological constant
\begin{eqnarray}
\langle |\Phi_\ast|^2 \rangle = \frac{\xi_2}{2 (\xi_2^2 + 4 \lambda \xi_1^2)} M_{Pl}^2 \equiv \frac{v^2}{2},  
\qquad
\Lambda = \frac{\lambda}{4 (\xi_2^2 + 4 \lambda \xi_1^2)} M_{Pl}^2,
\label{Higgs-min}  
\end{eqnarray}
where $v \approx 250GeV$. 

The first equality in Eq. (\ref{Higgs-min}) yields a relation
\begin{eqnarray}
\frac{\xi_2}{\xi_2^2 + 4 \lambda \xi_1^2} = \left(\frac{v}{M_{Pl}} \right)^2 \simeq 10^{-32}.
\label{Higgs-v}  
\end{eqnarray}
Next, using (\ref{Higgs-v}), the second equality in Eq. (\ref{Higgs-min}) produces a value of the cosmological
constant 
\begin{eqnarray}
\Lambda = \frac{\lambda}{4 \xi_2} v^2.
\label{Higgs-cc}  
\end{eqnarray}

A key observation here is that the parameters $\xi_1$ and $\xi_2$ are not limited to satisfy the various 
experimental constraints owing to the absence of the Einstein-Hilbert term in (\ref{Reduced-Lag}) \cite{Stelle}. 
Though we can choose any values of them, an interesting choice for $\xi_2$ might be $\xi_2 \simeq 10^{5-6}$
which comes from the Higgs inflation \cite{Bezrukov}. In any case, assuming that $\lambda \simeq 1,
\xi_2 \ll \xi_1$, we choose $\xi_1$ and $\xi_2$ to satisfy
\begin{eqnarray}
\frac{\xi_2}{\xi_1^2} \simeq \left(\frac{v}{M_{Pl}} \right)^2 \simeq 10^{-32}.
\label{Higgs-condition}  
\end{eqnarray}
Since $\frac{1}{\xi_1^2}$ corresponds to the gravitational coupling, Eq. (\ref{Higgs-condition}) means a very
weak coupling constant. Furthermore, as seen in Eq. (\ref{Higgs-cc}), it is true that the cosmological constant
is not so small compared to the magnitude that the cosmological observation implies, but it is very small
compared to the Planck mass squared and becomes smaller for the larger $\xi_2$.   

To close this section, it is of interest to imagine that the Higgs potential entirely comes from the gravity
by putting $\lambda = 0$. Under such a situation, the Higgs potential in (\ref{Higgs-pot}) is a perfect square
with a positive coefficient. This form of the Higgs potential implies two important facts; the automatic stability 
of the ground state and no cosmological constant at the minimum. In addition, this potential has emerged
from the requirement that the Einstein-Hibert term should appear. Otherwise, we could never have general 
relativity at low energies, which is against our world. To put it differently, the appearance of general relativity 
at low energies naturally leads to spontaneous symmetry breaking of the gauge symmetry to occur.  This situation 
should be contrasted to the conventional situation in the standard model: In the standard model, the
renormalizability of the theory requires that the Higgs potential takes a rather simple form
\begin{eqnarray}
V (\Phi) = m^2 |\Phi|^2 + \lambda |\Phi|^4,
\label{Higgs-SM}  
\end{eqnarray}
up to radiative corrections. For spontaneous symmetry breaking to occur, the renormalized value of the parameter
$m^2$ should be negative. But even the qualitative prediction that the symmetry is broken is not a prediction 
of the model. The parameter $m^2$ could have either sign; there is no logic that prefers one sign to the other. 
On the other hand, in our theory, the existence of general relativity at low energies naturally leads to
the spontaneous symmetry breakdown of the gauge symmetry.

\section{Derivation from BRST formalism}

We wish to understand the quantum equivalence between (\ref{Reduced-Lag}) and (\ref{F-Lag-star}).
In deriving (\ref{F-Lag-star}) in the Einstein frame from (\ref{Reduced-Lag}) in the Jordan frame,
we have heavily used the change of variables in Eq. (\ref{E-Weyl gauge transf}). Usually in quantum
field theory, the change of variables does not modify the physical content, but there has been
a prolonged controversy about the quantum equivalence of the theory between the two frames \cite{Kamenshchik}.
  
In this section, to clarify this issue, we wish to derive (\ref{F-Lag-star}) in the Einstein frame by beginning with 
(\ref{Reduced-Lag}) in the Jordan frame through a different but more sound formalism, that is, the
BRST formalism. The key idea is to show that the Lagrangian density in the Jordan frame is 
gauge-equivalent to that in the Einstein frame by taking a suitable gauge fixing condition for the Weyl
symmetry.

To do that, we fix the Weyl symmetry in such a way that a gauge fixing condition 
breaks only the Weyl invariance but leaves the general coordinate invariance unbroken. Then, a suitable gauge 
condition is 
\begin{eqnarray}
\tilde R + a |\Phi|^2 = b,
\label{general-GF}  
\end{eqnarray}
where $a, b$ are constants. This gauge choice certainly breaks only the Weyl invariance 
since the LHS has a Weyl weight $-2$ whereas the RHS does a vanishing Weyl weight, and the both sides 
are invariant under the general coordinate transformation. Incidentally, we cannot put $b = 0$ (``Landau
gauge'') since the quantities on the LHS of the gauge condition (\ref{general-GF}) transform 
only by an overall scale factor $\Omega^{-2}(x)$ under the Weyl gauge transformation.

Let us recall that the BRST transformation for the Weyl symmetry is given by \cite{Oda-E, Oda-S}
\begin{eqnarray}
\delta_B g_{\mu\nu} &=& 2 c g_{\mu\nu}, \qquad
\delta_B \sqrt{-g} = 4 c \sqrt{-g}, \qquad
\delta_B \tilde R = - 2 c \tilde R,
\nonumber\\
\delta_B \Phi &=& - c \Phi, \qquad \delta_B \bar c = i B, \qquad \delta_B c = \delta_B B = 0.
\label{BRST}  
\end{eqnarray}
Then, we find that a Lagrangian density for the gauge condition and the FP ghost reads \cite{Kugo} 
\begin{eqnarray}
{\cal{L}}_{GF+FP} &=& - i \delta_B \left[ \sqrt{-g} \, \bar c \left( \tilde R + a |\Phi|^2 - b 
+ \frac{\alpha}{2} B \right) \right]
\nonumber\\
&=& \sqrt{-g} \left[ \hat B \left( \tilde R + a |\Phi|^2 - b \right) + \frac{\alpha}{2} \hat B^2 
- 2 i b \bar c c \right]
\nonumber\\
&=& \sqrt{-g} \left[ - \frac{1}{2 \alpha} \left( \tilde R + a |\Phi|^2 - b \right)^2 - 2 i b \bar c c \right]
\nonumber\\
&=& - \sqrt{-g} \frac{1}{2 \alpha} \left( \tilde R + a |\Phi|^2 - b \right)^2 
\nonumber\\
&-& i \hbar \delta^4 (0) \log \left(b \sqrt{- g(x)} \right),
\label{GF-FP}  
\end{eqnarray}
where we have defined $\hat B \equiv B + 2 i \bar c c$, we performed the path integral over the 
auxiliary field $\hat B$, and in the last step we have done the integration over the FP ghosts 
\cite{Oda-E, Oda-S}. The last term proportional to $\delta^4 (0)$ has also appeared in Ref. \cite{Hayashi},
which we can neglect when we use the dimensional regularization.

Now, adding the gauge-fixing term (\ref{GF-FP}) to the Lagrangian density (\ref{Reduced-Lag}) 
in the Jordan frame, we can obtain a gauge-fixed and BRST-invariant Lagrangian density given by
\begin{eqnarray}
{\cal{L}} &=& \sqrt{-g} \Biggl[  \Bigl( \xi_1^2 - \frac{1}{2 \alpha} \Bigr) \tilde R^2 
+ \Bigl( \xi_2 - \frac{a}{\alpha} \Bigr) \tilde R |\Phi|^2 
- \Bigl( \lambda + \frac{a^2}{2 \alpha} \Bigr) |\Phi|^4
\nonumber\\
&-& \frac{1}{4} H_{\mu\nu} H^{\mu\nu} - |D_\mu \Phi|^2 - \frac{b^2}{2 \alpha} 
+ \frac{b}{\alpha} \tilde R + \frac{a b}{\alpha} |\Phi|^2  \Biggr].
\label{BRST-Lag}  
\end{eqnarray}
It turns out that this Lagrangian density (\ref{BRST-Lag}) coincides with the Lagrangian density 
(\ref{F-Lag-star}) in the Einstein frame when the parameters satisfy the following relations
\begin{eqnarray}
\alpha = \frac{1}{2 \xi_1^2}, \qquad  a = \frac{\xi_2}{2 \xi_1^2}, \qquad
b = \frac{1}{4 \xi_1^2} M_{Pl}^2.
\label{parameter}  
\end{eqnarray}

Actually, under the conditions (\ref{parameter}), Eq. (\ref{BRST-Lag}) is reduced to (\ref{F-Lag-star})
(without the symbol $\ast$ which simply means the Einstein frame, but is irrelevant to the present 
context) as follows:
\begin{eqnarray}
{\cal{L}} &=& \sqrt{-g} \Biggl[  \frac{M_{Pl}^2}{2} \tilde R - \frac{1}{4} H_{\mu\nu} H^{\mu\nu}
- |D_\mu \Phi|^2 
\nonumber\\
&-& \frac{1}{16 \xi_1^2} M_{Pl}^4 + \frac{\xi_2}{4 \xi_1^2} M_{Pl}^2 |\Phi|^2 
- \Bigl( \lambda + \frac{\xi_2^2}{4 \xi_1^2} \Bigr) |\Phi|^4  \Biggr]
\nonumber\\
&=& \sqrt{-g} \Biggl[ \frac{M_{Pl}^2}{2} R - \frac{1}{2} m_S^2 S_\mu S^\mu - \frac{1}{4} H_{\mu\nu} H^{\mu\nu}
- |D_\mu \Phi|^2 
\nonumber\\
&-& \lambda  |\Phi|^4 - \frac{1}{4 \xi_1^2} \left( \xi_2 |\Phi|^2 - \frac{M_{Pl}^2}{2} \right)^2 \Biggr].
\label{Desired-Lag}  
\end{eqnarray}
Thus, we have derived the Lagrangian density in the Einstein frame by starting with that in the Jordan frame
in the framework of the BRST formalism. The approach based on the BRST formalism is free from the problem
associated with the functional measure and provides a rather reliable method which demonstrates the equivalence
between the Jordan frame and the Einstein frame at the quantum level.

\section{Conclusion}

In this article, we have investigated a possibility that the Higgs potential is generated from the gravity.
To make this idea be more realistic, we have to solve two problems, one of which is related to the presence
of a massless dilaton and the other is the quantum equivalence of the theory between the Jordan frame
and the Einstein frame. 

We have solved the former problem by extending a global scale invariance to a local scale one where 
it turns out that the Weyl geometry provides a natural arena for formulating the local scale invariance
as the Weyl gauge invariance. In our theory, the Weyl gauge field becomes massive by eating the massless
dilaton and its magnitude of the mass is of order of the Planck mass with the Abelian coupling constant
$f \simeq 1$, so the Weyl gauge field might be a candidate for dark matter.  
 
As a resolution of the latter problem, an ongoing debate on the equivalence of the theory between the
Jordan frame and the Einstein frame, we have used the BRST formalism which does not depend on
the definition of the functional measure, and we have shown that up to a factor $\delta^4 (0)$ which can be
ignored in the dimensional regularization procedure, the theory in the both frames is gauge-equivalent. 

It is remakable that an $R^2$ term with the nonminimal coupling term $R |\Phi|^2$ gives us a Higgs
potential of a perfect square, by which the problem of the negative tachyonic mass in the Higgs potential
and the cosmological constant problem are solved. Boldly speaking, a complete resolution of the origin
of the Higgs potential in the standard model amounts to the problem of why the bare quartic interaction
$\lambda |\Phi|^4$ is zero.     

Finally, we wish to mention some future problems. It is straightforward to extend the present theory
to the standard model and the grand unified models by extending the gauge group and the definition
of the covariant derivatives. Another interesting question is to verify that the massive Weyl gauge
field could be really a candidate of dark matter by the explicit calculation. The other problem
is to introduce a manifestly scale-invariant regularization technique to treat with the Weyl gauge
symmetry without anomalies. We wish to return these problems in future.

\begin{flushleft}
{\bf Acknowledgements}
\end{flushleft}

We would like to thank T. Kugo for valuable discussions and careful reading of this manuscript.


\end{document}